# Expectile-based hydrological modelling for uncertainty estimation: Life after mean

Hristos Tyralis[1,2,3], Georgia Papacharalampous[3], Sina Khatami[4,5]

[1]Construction Agency, Hellenic Air Force, Mesogion Avenue 227–231, 15 561 Cholargos, Greece (montchrister@gmail.com, https://orcid.org/0000-0002-8932-4997)

[2]Department of Water Resources and Environmental Engineering, School of Civil Engineering, National Technical University of Athens, Iroon Polytechniou 5, 157 80 Zografou, Greece (hristos@itia.ntua.gr)

[3]Department of Topography, School of Rural, Surveying and Geoinformatics Engineering, National Technical University of Athens, Iroon Polytechniou 5, 157 80 Zografou, Greece (papacharalampous.georgia@gmail.com, https://orcid.org/0000-0001-5446-954X)

[4]Department of Earth Sciences, Uppsala University, Villavägen 16, 75236 Uppsala, Sweden (sina.khatami@geo.uu.se, https://orcid.org/0000-0003-1149-5080)

[5]Department of Infrastructure Engineering, University of Melbourne, Parkville, Victoria, Australia

**Abstract**: Predictions of hydrological models should be probabilistic in nature. Our aim is to introduce a method that estimates directly the uncertainty of hydrological simulations using expectiles, thus complementing previous quantile-based direct approaches as well as generalizing mean-based approaches. Expectiles are new risk measures in hydrology. Compared to quantiles that use information of the frequency of process realizations over a specified value, expectiles use additional information of the magnitude of the exceedances over the specified value. Expectiles are least square analogues of quantiles and can characterize the probability distribution in much the same way as quantiles do. Moreover, the mean of the probability distribution is the special case of the expectile at level 0.5. To this end, we propose calibrating hydrological models using the expectile loss function, which is strictly consistent for expectiles. We apply our method to 511 basins in contiguous US and deliver predictive expectiles of hydrological simulations with the GR4J, GR5J and GR6J hydrological models at expectile levels 0.5, 0.9, 0.95 and 0.975. An honest assessment empirically proves that the GR6J model outperforms the other two models at all expectile levels. Great opportunities are offered for moving beyond the mean in hydrological modelling by simply adjusting the objective function.

# 1. Introduction

Properties of hydrological variables can be characterised by functionals of their probability distribution, e.g. quantiles. A representative example is rainfall intensity-duration-frequency curves, in which the intensity for specified rainfall durations and frequencies is a quantile of the rainfall distribution. Characterising the uncertainties of hydrological model simulations and predictions in general (Beven and Binley 1992, Beven and Young 2013, Khatami et al. 2019, Todini 2007) by estimating quantiles of their probability distributions is another relevant example. In particular, evaluating hydrological modelling uncertainties including prediction uncertainties has been recognized as an important problem to solve (Blöschl et al. 2019), which has been extensively studied (Beven 2009, Montanari 2011, Solomatine and Wagener 2011). Modelling approaches that estimate predictive uncertainty can be classified as follows:

a. Estimation directly through the model, using Bayesian techniques, Monte Carlo schemes or less frequently suitable loss functions, e.g. (Althoff et al. 2021, Ammann et al. 2019, Beven and Binley 1992, Evin et al. 2014, Hernández-López and Francés 2017, Kuczera et al. 2006, Tyralis and Papacharalampous 2021b).

b. Estimation by post-processing residual errors using a second model, see Supplementary material.

Predictions of hydrological models should be distributional (probabilistic or possibilistic) in nature, taking the form of probability distributions over predicted variables (e.g. as done in many of above references); however, many practical situations require point predictions (Gneiting 2011) (see also many of the above references). It is common practice to deliver point predictions close to the materialization of the hydrological variable of interest (e.g. streamflow). To do so, one can set the objective (or loss, score etc.) function to calibrate the hydrological model (Solomatine and Wagener 2011), e.g. the squared error loss function or a similar one (e.g. the Nash - Sutcliffe efficiency (Nash and Sutcliffe 1970), the absolute error function and more).

Point predictions close to the observations are useful and in fact characterize the vast majority of hydrological modelling studies. However, quantiles (which are also point predictions) of the predictive distribution may also be of interest. For instance, quantile predictions at a dense grid of quantile levels can characterize the predictive distribution of hydrological variables (Brehmer and Gneiting 2021). It may also be of interest a specific quantile of the predictive distribution, e.g. a quantile at level 97.5%, for flood design purposes (that is practically the essence of the return period concept for flood prediction).

To this end, Tyralis and Papacharalampous (2021b) proposed calibration of hydrological models using the quantile loss function (Koenker and Bassett Jr 1978) to provide point predictions of specified quantiles of the predictive distribution. Quantile-based hydrological modelling and Bayesian frameworks, for characterizing predictive uncertainty share the property that they can directly provide predictive distributions (those approaches belong to the class of modelling approaches that estimate predictive uncertainty using Bayesian techniques or suitable loss functions, as discussed earlier in this Section). However, they are dissimilar with respect to other properties, as explained in Tyralis and Papacharalampous (2021b) and repeated in Section 5 for the sake of completeness.

Expectiles, introduced by Newey and Powell (1987) (see also early developments in Aigner et al. (1976)), are least squares analogues of quantiles in the sense that they are generalizations of the mean, in the same way that quantiles are generalizations of the median (Daouia et al. 2018). Therefore, they can characterize the predictive distribution in much the same way as quantiles do. A major advantage of estimating expectiles compared to quantiles is that available information is used more efficiently. In particular, estimation of quantiles requires knowledge of the location of materializations (realizations) of the process (i.e. whether materializations are below or above the predictors), but estimation of expectiles exploits an additional information content of model residuals, i.e. the distance between each modelled value and its corresponding observation (Daouia et al. 2018). The concept of expectiles is controversial in the statistical literature regarding mostly issues of interpretability (Koenker 2013) and their use remains limited in hydrological applications. However, expectiles application as a risk measure is steadily growing, albeit at lower rates compared to quantiles (Waltrup et al. 2015). Section 2.3 aims to review the basics of expectiles and demonstrate their utility for the hydrological community.

Our aim is to introduce a method that estimates the uncertainty of hydrological simulations using expectiles, given their potential as an uncertainty (risk) measure. As a proof of concept, we showcase numerical examples for quantifying predictive expectiles of simulated streamflow at the daily time scale. To do so, we propose calibrating a hydrological model using the expectile loss function and we built on the framework proposed in Tyralis and Papacharalampous (2021b), which provided directly predictive quantiles for characterizing predictive uncertainty in hydrological modelling. Our study:

(a) Complements the literature of uncertainty quantification in hydrological modelling using direct methods.

(b) Generalizes approaches that are based on calibrating hydrological models with squared error loss function, which is a special case of the expectile loss function.

The remainder of the manuscript is structured as follows. Section 2 provides an explanation of expectiles and their theoretical properties as well as an outline of concepts related to the expectile loss function. Since we are aware of the limited use of expectiles in hydrology, as well as the extensive discussion on their utility in the general literature, we emphasize reasons to use them, based on a concise summary of the statistical literature. The GR4J, GR5J and GR6J daily lumped hydrological models that were used in Coron et al. (2017) to demonstrate our method are described briefly in Section 3.1. We apply our method to 511 basins in the contiguous US, using the three hydrological models, and summarize the key components of the proposed approach in Section 3.3. The results including comparison of the performance of the models are presented in Section 4. In Section 5, we further discuss the differences between our proposed expectile-based approach and the existing methods for evaluating modelling uncertainties. The paper closes with the Conclusions (Section 6).

## 2. Theoretical background

Here, we present material related to the expectile loss function as well as the concept of expectiles. The relevant concepts of the quantile loss function and the quantile are presented, and their similarities and differences are discussed. We close the Section by presenting the hydrological models used in the study.

## 2.1 Quantiles and quantile loss function

Let $\underline{x}$ be a real-valued random variable with distribution function $F$ defined by:

$$F(x) := P(\underline{x} \leq x) \quad (1)$$

The $a^{th}$ quantile of $\underline{x}$ is defined by

$$q_a := \inf\{x : F(x) \geq a\} \quad (2)$$

By inverting eq. (2), one can obtain the following equation for a general class of probability distribution functions

$$a = E[\mathbb{1}(\underline{x} \leq q_a)] \quad (3)$$

where $\mathbb{1}(\cdot)$ denotes the indicator function. The sample analogue of eq. (3) implies that the quantile level $a$ corresponds to the frequency (recall the definition of intensity-duration-frequency curves in Introduction Section 1) of realizations of $\underline{x}$ that are below $q_a$. For instance, when $a = ½$, half of realizations are expected to be below the median $q_{1/2}$ of $F$.

In regression modelling, the $a^{th}$ quantile of the predictive distribution is obtained by minimizing the expected value $E[L_1(r; \underline{x}, a)]$ with respect to $r$, where $L_1(r; x, a)$ is defined by eq. (4).

$$L_1(r; x, a) := (r - x)(\mathbb{1}(x \leq r) - a) \quad (4)$$

Here, $x$ is the materialization of the variable $\underline{x}$, $a$ is the quantile level of interest, and $r$ is the respective predictive quantile. The idea of using the quantile loss function was elaborated by Koenker and Bassett Jr (1978). When $a = ½$, eq. (4) reduces to

$$L_1(r; x, 1/2) = |r - x|/2 \quad (5)$$

which is half the absolute error function, and its minimization yields the median $q_{1/2}$. It is then obvious that quantiles are generalizations of the median (Koenker 2005, Koenker 2017).

## 2.2 Expectiles and expectile loss function

Similar to eq. (3), the expectile $e_\tau$ at the expectile level $\tau$ can be defined by inverting the following equation (Daouia et al. 2018):

$$\tau = E[|x - e_\tau|\mathbb{1}(\underline{x} \leq e_\tau)] / E[|x - e_\tau|)] \quad (6)$$

A definition of expectiles has been given by Newey and Powell (1987):

$$e_\tau := \mathrm{argmin}_{\theta \in R} \, E[L_2(\theta; \underline{x}, \tau) - L_2(0; \underline{x}, \tau)] \quad (7)$$

where

$$L_2(r; x, \tau) := (r - x)^2 \, |\mathbb{1}(x \leq r) - \tau| \tag{8}$$

Comparing eqs. (3) and (6), one sees that $q_a$ is the point below which $100a\%$ of the mass of $\underline{x}$ lies, while $e_\tau$ is such that the mean distance from all $\underline{x}$ below $e_\tau$ is $100\tau\%$ of the mean distance between $\underline{x}$ and $e_\tau$ (Daouia et al. 2018). Consequently, an intuitive interpretation of the expectile is that it is similar to the quantile when replacing frequency with distance.

Predictive expectiles in regression settings can be obtained similarly to predictive quantiles (see Section 2.1). In particular, the $\tau^{\text{th}}$ expectile of the predictive distribution is obtained by minimizing the expected value $\mathrm{E}[L_2(r; \underline{x}, \tau)]$ with respect to $r$, where $L_2(r; x, \tau)$ is defined by eq. (8) (Newey and Powell 1987). When $\tau = \frac{1}{2}$, eq. (8), reduces to

$$L_2(r; x, 1/2) := ((r - x)^2)/2 \tag{9}$$

which is half the squared error function, and its minimization yields the mean $e_{1/2}$. It is then obvious that expectiles are generalizations of the mean (Newey and Powell 1987) and the expectile loss function is the least-squares analogue of the quantile loss function.

Figure 1 presents the quantile and expectile scores when $x = 0$ materializes, for varying predictive quantiles and expectiles $r$ at the quantile levels $a \in \{0.05, 0.25, 0.75, 0.95\}$ and expectile levels $\tau \in \{0.05, 0.25, 0.75, 0.95\}$. Both loss functions are minimized when $r = x$, with minimum value equal to zero. Furthermore, both loss functions are asymmetric (for instance see the asymmetry of the expectile loss function at level 0.95 as depicted with the blue solid line); therefore, the degree of penalization depends on the location (below or above) of $r$ compared to $x$. The asymmetry is a result of assigning different weights to the error $r - x$ depending on its sign and allows estimating quantiles or expectiles at levels different from $\frac{1}{2}$. At level $\frac{1}{2}$, both functions become symmetric; see eqs. (5) and (9). A significant difference is that the expectile loss function assigns disproportionally higher scores to values that are distant compared $x$.

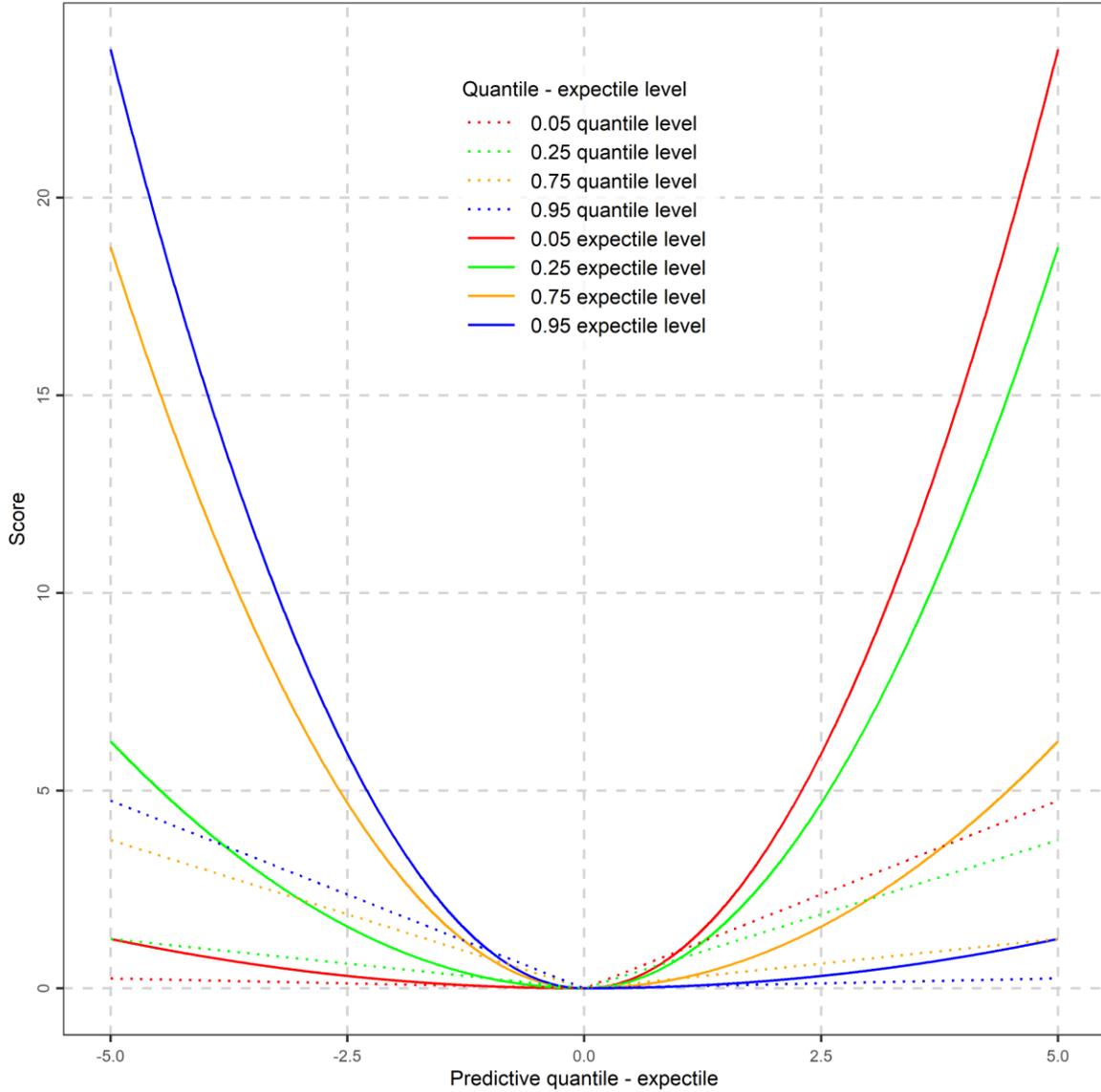

Figure 1. Illustration of the quantile and expectile scores at the quantile levels $a \in \{0.05, 0.25, 0.75, 0.95\}$ and expectile levels $\tau \in \{0.05, 0.25, 0.75, 0.95\}$ respectively when $x = 0$ materializes and for varying predictive quantiles and expectiles $r$ (see eqs. (4) and (8)).

## 2.3 Properties of expectiles and of the expectile loss function

### 2.3.1 Illustrative arithmetic example

We start with an illustrative arithmetic example on the differences between quantiles and expectiles. The example is especially relevant to statistical properties of the daily streamflow related to extreme magnitudes. In particular, empirical studies (Blum et al. 2017) have shown that distributions suitable for modelling daily streamflow belong to the domain of attraction of the Fréchet distribution, when streamflow daily maxima are sampled at annual blocks. The Fréchet distribution is a special case of the Generalized Extreme Value (GEV) distribution with shape parameter $k > 0$, while large-scale empirical studies have shown that the Fréchet distribution is suitable to model annual daily streamflow maxima (Tyralis et al. 2019c).

Figure 2 presents a simulation of the Generalized Pareto (GP) distribution with location parameter $\mu = 0$, scale parameter $\sigma = 1$ and shape parameter $\xi = 0.2$. The GP distribution belongs to the domain of attraction of the Fréchet distribution for $x > \mu$, with parameters $\xi$ and $k$ being equal.

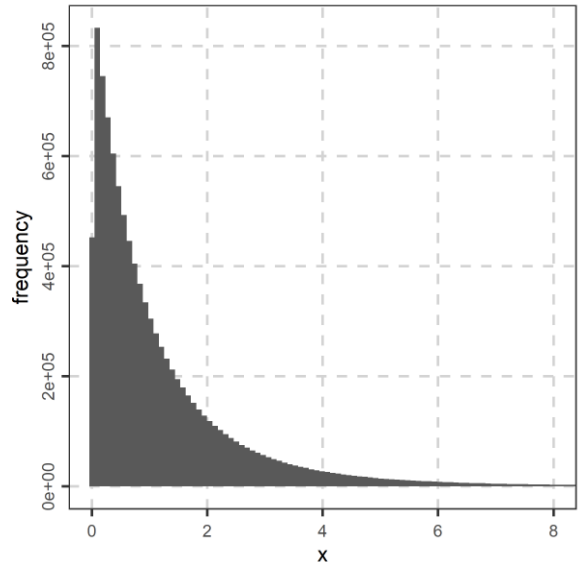

Figure 2. Histogram (10 000 000 samples) (truncated at 8 in the x-axis for visualization reasons) produced by the simulation of a generalized Pareto distribution with $\mu = 0$, $\sigma = 1$ and $\xi = 0.2$.

At level 0.975, the sample quantile of the simulated $\underline{x}$ is equal to 5.46, while the sample expectile is equal to 4.66. Now we add the quantity 0.1 at all samples higher than 5.46. The new sample quantile at level 0.975 remained 5.46; however, the new sample expectile increased to 4.70. New sample quantiles at lower levels have not changed either, while new sample expectiles increased at all levels. Please recall here Section 2.2, in which it was shown that quantiles are insensitive to distances on contrary to expectiles, while they depend only on the mass below or above the quantile at the level of interest. With the above example, it was shown that quantiles are insensitive to the magnitude of extreme events, while they depend only on their frequency. Therefore, in empirical settings, two distributions with different tail behaviours can have identical quantiles (Kuan et al. 2009).

Furthermore, the severity of the exceedances may be of interest, e.g. when examining floods, in which case quantiles might be less appropriate risk measures (Kuan et al. 2009, Taylor 2008). On the contrary, expectiles are more tail sensitive (Kuan et al. 2009).

Hydrologists are used to think in terms of the concept of return period to characterize risks. Returning to the artificial example of the present Section, it is obvious that the increased severity of the events did not change the return period of 40 years (which corresponds to a quantile level 0.975), that characterizes the sample quantile value 5.46. After defining a new type of return period that is associated to expectile levels, we compute that this new type of return period has decreased to 39 years from 40 years for the value of the expectile 4.66, as one should ask from her/his risk measure.

According to conventions and engineering experience, structures are designed after specifying return periods and estimating related variable values that should not be exceeded beyond the frequency that corresponds to the specified return period. Designing with frequency principles that are directly related to quantiles, may have some disadvantages, albeit switching to a new risk measure would require to change some conventions (e.g. specifying values for the new type of return period that may be different compared to values of frequency-based return periods). The practice of searching for appropriate risk measures is common in banking regulation; see e.g. the discussion related to changes of the Basel protocol (Ehm et al. 2016, Chen 2018).

As clarified in the next Sections 2.3.2 and 2.3.3, expectiles have some favourable theoretical properties compared to existing risk measures. Among those properties the most discussed seem to be that they are the only elicitable and coherent risk measures among *M*-quantiles (see also related definitions and references in the next Sections).

*2.3.2 Estimation of expectiles, loss functions, proper scoring rules, elicitability and consistency*

An expectile is an elicitable function of the predictive distribution, i.e. there exists a strictly consistent loss function for it (Gneiting 2011). If a modeller receives a directive to predict a statistical functional (e.g. a quantile or an expectile) of the probability distribution, then a loss function is strictly consistent for the statistical functional when its expectation is uniquely minimized when following the directive (Gneiting 2011). It is proved that the squared error loss function is strictly consistent for the mean functional, and the expectile loss function is strictly consistent for the expectile. Similarly, the absolute error function is strictly consistent for the median, and the quantile loss function is strictly consistent for the quantile (Gneiting 2011), while both quantile regression and expectile regression are classes of distributional regression (Kneib et al. 2021). It is important to use strictly consistent loss functions in point prediction; otherwise, predictive inference can be misguided (Gneiting 2011).

Scoring rules assess the quality of probabilistic predictions by delivering a score based on the prediction and the materialization of a process (Gneiting and Raftery 2007). "*A scoring rule is proper if truth telling is an optimal strategy in expectation*" (Brehmer and Gneiting 2020). Proper scoring rules encourage the modeller to make careful assessments and to be honest (Gneiting and Raftery 2007). The expectile loss function is consistent for the expectile relative to a class of distributions, therefore it induces a proper scoring rule (for details on the construction of the proper scoring rule see Ehm et al. 2016).

*2.3.3 Theoretical properties*

As mentioned, quantiles and expectiles are elicitable. Quantiles and expectiles are members of the class of *M*-quantiles (Breckling and Chambers 1988). In the literature of econometrics, quantiles and expectiles are viewed as risk measures. Among *M*-quantiles, the only coherent risk measures are expectiles (Bellini et al. 2014), since quantiles are not sub-additive (Taylor 2008). Furthermore, it has been proved that the only jointly coherent and elicitable risk measures are expectiles (Ziegel 2016) (see definitions of coherence for risk measures in Ziegel (2016)).

From a more practical point of view, expectiles are sensitive to the size of loss beyond a specified threshold; see also the illustrative example in Section 2.3.1 (Ziegel 2016). However, considering the size of loss in the estimation procedure comes at the cost of increased sensitivity to outliers. Relevant properties have been discussed in Koenker (2005 p.64) which argues that quantiles depend on local features of the distribution, whereas expectiles have a more global dependence. For instance, as it was shown in Section 2.3.1, altering the tail of the distribution did not change lower quantiles, while all expectiles increased.

The interpretability of expectiles compared to quantiles are discussed within the literature (Chen 2018, Kneib et al. 2013a,b, Koenker 2013, Kokic et al. 1997). Eilers (2013) advise against comparison of quantiles and expectiles magnitudes at the same level, albeit such comparisons are common in the literature (e.g. Abdous and Remillard 1995, Bellini and Di Bernardino 2017, Bellini et al. 2014). Furthermore, transformations between quantiles and expectiles also exist (Jones 1994). Regarding magnitudes, extreme expectiles are larger compared to extreme quantiles for very heavy-tailed distributions (i.e. ones with tail index higher than ½) while the opposite happens when the tail index is lower than ½ (Bellini et al. 2014).

Research has focused on how to obtain extreme expectiles or variants (i.e. expectiles at very large levels in which estimators are not robust) based on extreme value theory (Daouia et al. 2018, 2019, 2020, 2021, Mao et al. 2015), interpretability issues in general (Efron 1991, Ehm et al. 2016, Philipps 2021) or smoothing procedures and transforming expectiles to quantiles and, therefore, obtaining the full distribution (Schnabel and Eilers 2009). Similarly to quantile regression that is exploited in statistical and machine learning (Taylor 2000, Tyralis and Papacharalampous 2021a, Tyralis et al. 2019b), the same holds for expectile regression (Yin and Zou 2021).

## 3. Methods

### 3.1 Hydrological models

To assess our method, we used the Génie Rural (GR) GR4J, GR5J and GR6J daily lumped hydrological models, implemented in the `airGR` R package by Coron et al. (2017). These versions allow forcing calibration with a user-defined loss function. The GR hydrological models are widely implemented worldwide; therefore, a detailed description is not necessary (and out of the scope of our study). Their implementation in `R` allows reproducibility of the results. GR hydrological models have also been applied to present the concept of quantile-based hydrological modelling in Tyralis and Papacharalampous (2021b) and post-processing hydrological simulations in Papacharalampous et al. (2019) and Tyralis et al. (2019a). GR4J has four free parameters (Perrin et al. 2003), and complex GR5J and GR6J models have five and six parameters, respectively (Pushpalatha et al. 2011). The additional parameter of the GR5J (compared to GR4J) relates to inter-catchment water exchanges, while the two additional parameters of the GR6J model allow more flexibility for modelling the low flows (Coron et al. 2017). Models are calibrated based on the Michel's (1991) optimization algorithm.

### 3.2 Data

The hydrological models were implemented to simulate daily streamflow in 511 small- to medium-sized river basins of the CAMELS dataset (Addor et al. 2017a, b, Newman et al. 2014, 2015, 2017, Thornton et al. 2014) in the contiguous US (CONUS); see Figure 3. Daily time series of streamflow, precipitation and minimum and maximum temperatures are available for each river basin for the 34-year period of 1980–2013 (Thornton et al. 2014). The same period and river basins have been previously examined for quantifying predictive uncertainties (Papacharalampous et al. 2019, Tyralis and Papacharalampous 2021b, Tyralis et al. 2019a) and are again examined here for consistency reasons. We assumed mean daily temperatures for each river basin as the average of its minimum and maximum daily temperatures. Daily potential evapotranspiration was estimated by applying the Oudin's formula (Oudin et al. 2005) to the daily mean temperature time.

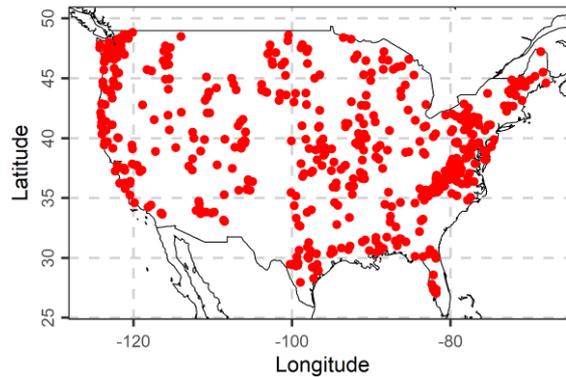

Figure 3. The 511 river basins over CONUS examined in the study.

## 3.3 Methodology outline

Here, we show how elements presented in Sections 2, 3.1 and 3.2 are combined in the context of computational experiments on streamflow simulations. The elements of the method are:

   i. Hydrological models: GR4J, GR5J and GR6J implemented in the `airGR` R package.
   ii. Model inputs: Daily precipitation and daily potential evapotranspiration.
   iii. Model output: Daily streamflow.
   iv. Objective function: Expectile loss function.
   v. Period of interest: 1980-2013, with 1980-1981 for model warm-up, 1982-1997 for model calibration, and 1998-2013 for model evaluation.
   vi. Spatial coverage: 511 river basins in CONUS.

Split-sample test (Biondi et al. 2012, Klemeš 1986) for each model run is conducted as:

   i. Warm up of the models over the 2-year 1980-1981.
   ii. Calibration of the models over the 16-year period 1982-1997 using the expectile loss function $L_2(r; x, \tau)$ at expectile levels $\tau \in \{0.5, 0.9, 0.95, 0.975\}$. That equates to 12 parameter sets (4 loss functions × 3 models), consequently 12 distinct model runs for each catchment.

iii. Evaluation of the streamflow simulation for the 16-year period 1998-2013 for each model run.

The computations and visualizations were conducted in R Programming Language using the contributed packages presented in the Supplementary material.

## 4. Results

First, we present a visual illustration of the expectile-based model outputs in Figure 4, for an arbitrary basin over 2012. The hydrological models were calibrated against expectile loss at levels 0.5 and 0.975, in which case they simulate expectiles at respective levels. They were also calibrated with the quantile loss function at equal levels. Results of the GR5J and GR6J model are omitted in Figure 4 for reasons of improved visualization. Furthermore, results for the remaining two hydrological models would lead to the same conclusions regarding the explanation of Figure 4.

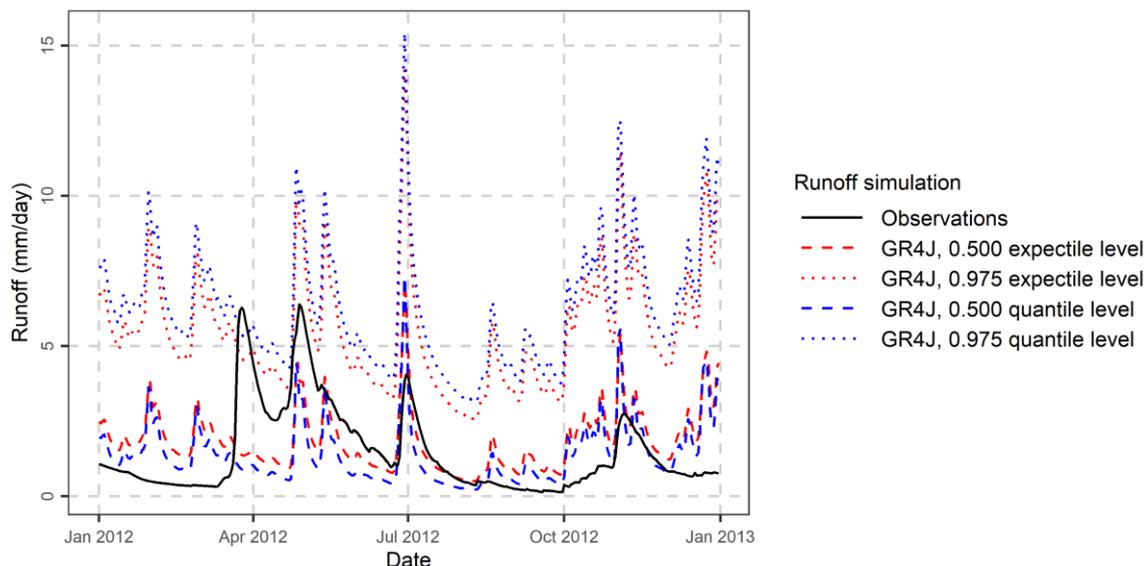

Figure 4. Illustration of observed streamflow, quantiles and expectiles simulations by the GR4J model at the levels $a \in \{0.5, 0.975\}$ for one-year period at an arbitrary river basin.

Recall that the expectile loss function at level 0.5 is equivalent to the squared error loss, while the quantile loss function at level 0.5 is equivalent to the absolute error loss. Therefore, simulations of three models at level 0.5 are close to the observations. On the other hand, simulations at level 0.975 are in general higher compared to the observations. Simulated expectiles at level 0.975 are in general lower compared to respective quantiles. This result is expected according to the theoretical properties of the expectiles (see

Section 2.3.3), as streamflow is characterized by a tail index lower than ½.

As expectile scores are not scaled, it is difficult to directly assess the performance of multiple models at multiple catchments. To overcome this, we used relative scores (Hyndman and Koehler 2006). Here we set the simpler model as a benchmark model, and compare other models against the benchmark. We define the relative score in eq. (10).

$$SCORE_{rel} := (SCORE_{bench} - SCORE_{model})/SCORE_{bench} \qquad (10)$$

Interpretation of the estimated expectiles in Figure 4 can be based on their properties, already presented in Sections 2.2 and 2.3. For instance the predictive expectile at level 0.975 is such that the mean distance from all daily streamflows below the expectile is 97.5% of the mean distance between daily streamflows and the respective expectile. On the contrary the predictive quantile at level 0.975 is such that daily streamflows fall with a frequency 97.5% under the respective predictive quantile.

We set GR4J as the benchmark model and compared GR5J and GR6J against it at expectiles levels 0.5, 0.9, 0.95 and 0.975. Figure 5 presents the relative scores at all basins and all levels. On median, the overall performance of GR5J across all model runs is 1.49% lower than GR4J while GR6J is higher by 3.73%, given the expectile loss function and assumptions of our experiments.

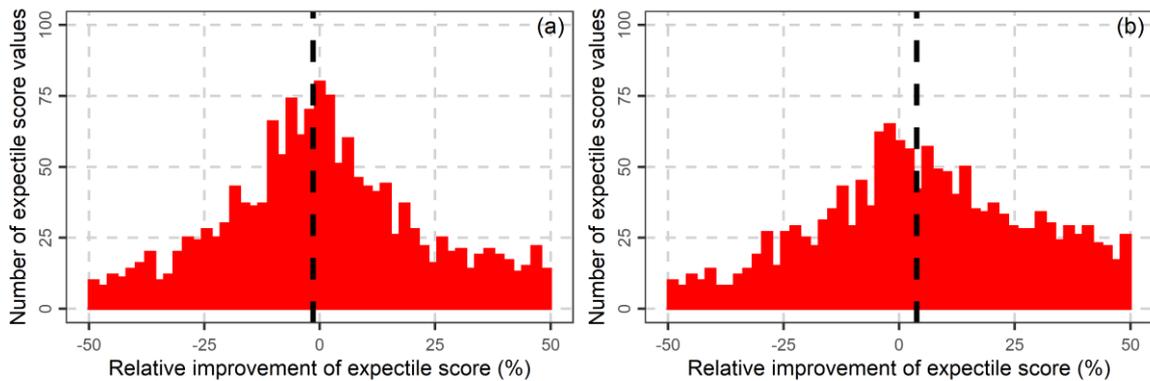

Figure 5. Histograms of the relative improvements (with red colour) and the median values of these relative improvements (with black dashed lines) against the performance of the GR4J hydrological model in terms of expectile score, as computed for all the 511 river basins and the 0.5, 0.9, 0.95, 0.975 expectile levels, for the (a) GR5J and (b) GR6J hydrological models. Truncation at −50% and 50% has been applied for illustration purposes.

Also, to compare models at each expectile level, we presented a summary of relevant scores for all basins at each expectile level in Figure 6. The more complex models GR5J and GR6J improve approximately 10% over the GR4J at the expectile level 0.5.

Improvements are less pronounced for the GR6J model at higher expectile levels, while the GR5J degrades with respect to the GR4J model.

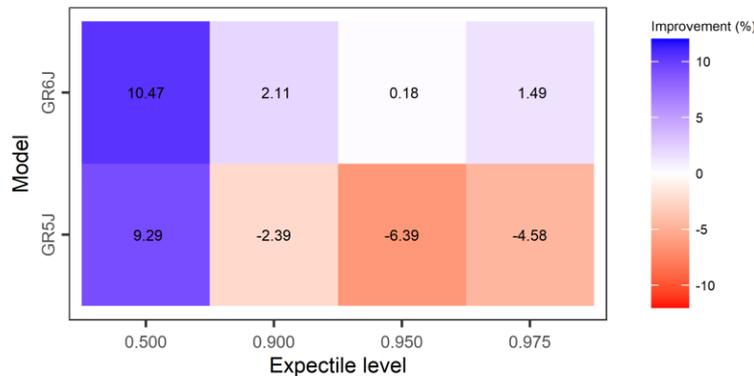

Figure 6. Heatmap of the median of the relative improvements summarizing the results for the 511 river basins for the performance of the GR5J and GR6J models against the performance of the GR4J hydrological model in terms of expectile scores.

Since expectile scores are not intuitive and scaled, we also examined the hydrological models' properties with respect to a component of the expectile score, i.e. the respective expectile level. In particular, we computed the sample expectile level based on eq. (6) and respective results are summarized for all basins in Figure 7. Recall that expectile levels are the equivalent of quantile levels, when accounting for the distance. For instance, when calibrating with the squared error loss function, the median of the expectile level estimates at each basin should be 0.500 for ideal simulations at all basins (0 model residuals). GR6J is uniformly better compared to the GR5J, which in turn is uniformly better compared to the GR4J model at all expectile levels.

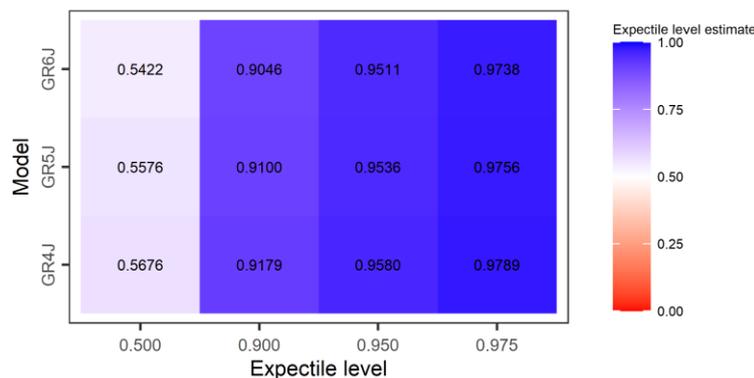

Figure 7. Heatmap of the median of the expectile level estimates of the predictions issued by the GR4J, GR5J and GR6J hydrological models at varying expectile levels summarizing the results for the 511 river basins.

Results of Figures 6 and 7 may seem contradicting but in fact they are not comparable. In particular, recall from Section 2.3.2 that if a modeller receives a directive to predict a statistical functional of the probability distribution, then a loss function is strictly consistent for the statistical functional when its expectation is uniquely minimized when following the directive. Furthermore, it is important to use strictly consistent loss functions in point prediction; otherwise, predictive inference can be misguided. The expectile loss function is strictly consistent for the expectile. However, the loss function that issues a penalty based on the sample expectile level (e.g. the loss function used in Figure 7) is not strictly consistent for the expectile.

## 5. Some more remarks

To provide point predictions of specified quantiles of the predictive distribution, Tyralis and Papacharalampous (2021b) proposed calibration of hydrological models using the quantile loss function (Koenker and Bassett Jr 1978). In Tyralis and Papacharalampous (2021b), the differences between those two approaches were discussed based on the theoretical argumentation by Waldmann (2018). Expectiles are least-squared analogues of quantiles; therefore, they share similar properties to quantiles when compared to Bayesian techniques. Here, we repeat the discussion by Tyralis and Papacharalampous (2021b) and we cite differences between expectile regression and Bayesian techniques. In particular, expectile regression (and, therefore, our methodological approach as well) is appropriate in the following situations:

    a. When the interest lies on events far from the mean.

    b. When the distribution of the dependent variable is unknown.

    c. In the presence of heteroscedasticity.

    d. When fast estimation of functionals of the predictive distribution is needed. Expectile regression is faster compared to Bayesian techniques, since a simple optimization is needed.

These characteristics make it suitable to model daily streamflow processes, which are heteroscedastic while their distribution is unknown. Some undesirable properties of expectile regression compared to Bayesian methods are the following:

    a. Inference on the parameters (e.g. computation of confidence intervals and hypothesis testing) is complicated.

    b. Expectile crossing is a possibility.

    c. The full posterior conditional distribution is not available. It is possible to compute

multiple expectiles; however, these should first be transformed to quantiles, while expectiles at a dense grid are needed (Brehmer and Gneiting 2021). Consequently, multiple optimizations are needed to obtain expectiles at different levels, thus increasing computation times.

    d. Their existence imposes constraints on the tails heaviness (Daouia et al. 2021).

Differences between quantile and expectile regression, as well as differences between quantiles and expectiles as risk measures have already been discussed in Section 2; therefore, it is not necessary to repeat them here. We would like, however, to point out here again that both of them are risk measures. Specifically, by specifying a level $a$ for quantiles or $\tau$ for expectiles, $q_a$ is the point below which $100a\%$ of the mass of $\underline{x}$ lies, while $e_\tau$ is such that the mean distance from all $\underline{x}$ below $e_\tau$ is $100\tau\%$ of the mean distance between $\underline{x}$ and $e_\tau$. Thus, quantiles are connected to frequencies, while expectiles are also connected to the severity of losses leading to more coherent and realistic risk measures (Daouia et al. 2021). Regarding the sensitivity of expectiles to the sample size, comparative properties of the absolute and the squared error loss functions are transferable to a certain extent here, since it remains an open research question how these properties are transferable to the tails of the underlying distributions (i.e. at levels different to 0.5).

Moving from the concept of expectiles and focusing on the computational experiment, we have shown that, for the specific dataset and conditions, the QR6J model outperforms two other models in predicting expectiles at levels 0.5, 0.9, 0.95 and 0.975. The comparison is based on a consistent loss function and the magnitude of the dataset allows drawing reliable conclusions (see e.g. the discussions in Papacharalampous and Tyralis 2020, Papacharalampous et al. 2021, 2022). The results do not agree with those by Tyralis and Papacharalampous (2021b), where it was shown that none of the GR hydrological models outperforms the others when predicting quantiles. Perhaps, the additional complexity of GR models specialized them to simulate better the mean (i.e. using the squared error loss function and its variants), while this behaviour does not govern simulation of quantiles with the quantile loss function.

## 6. Conclusions

We introduced expectiles as a new approach for characterising uncertainty and risk-related measures in hydrology. They are least-squares analogues of quantiles and can summarize the predictive distribution of hydrological simulations in much the same way as quantiles do. They are more sensitive when modelling the tail of variables, due to exploiting distance-based information in addition to frequencies. To estimate expectiles of hydrological simulations, we proposed calibration of hydrological models with the expectile loss function. The expectile loss function is strictly consistent for expectile prediction.

As a proof of concept, we applied our method to simulate daily streamflow at expectile levels 0.5, 0.9, 0.95 and 0.975 at 511 basins in contiguous US. Simulations of expectiles were delivered at those levels. An honest assessment of the model performances was performed, in which it was empirically proved that the GR6J hydrological model performs better than the GR4J and GR5J hydrological models. This paper is a continuation of the quantile-based hydrological modelling paper by Tyralis and Papacharalampous (2021b), in which it was shown that hydrological models could simulate streamflow beyond its mean, after appropriate adjustments of the objective function. Those modelling techniques are a nice alternative to Bayesian techniques that are widely applied to estimate hydrological predictive uncertainty and prove that there is life after mean for hydrological models.

**Conflicts of interest:** The authors declare no conflict of interest.

**Acknowledgements:** We thank the Associate Editor and the reviewers, whose comments were important in improving the manuscript. SK has received funding by Mannerfelt fond, Ahlmanns fond, and the European Research Council (ERC) within the project HydroSocialExtremes: Uncovering the Mutual Shaping of Hydrological Extremes and Society, ERC Consolidator Grant No. 771678, H2020 Excellent Science. The funders did not interfere in the research design, data collection and analysis, and in the preparation of the manuscript.

**Author contributions:** HT and GP conceptualized the work and designed the experiments with input from SK. HT and GP performed the analyses and visualizations, and wrote the first draft, which was commented and enriched with new text, interpretations and discussions by SK.